%% file: main.tex
\documentclass[conference]{IEEEtran}
\IEEEoverridecommandlockouts
\IEEEaftertitletext{\vspace{-1.25\baselineskip}}

\ifCLASSOPTIONcompsoc
  \usepackage[nocompress]{cite}
\else
  \usepackage{cite}
\fi

\usepackage{amsmath,amssymb,amsfonts}
\usepackage{graphicx}
\usepackage{textcomp}
\usepackage[table]{xcolor}
\usepackage{varwidth}
\usepackage{afterpage}
\usepackage{babel}
\usepackage{collcell}
\usepackage{hhline}
\usepackage{multirow}
\usepackage{tikz}
\usepackage{qcircuit}
\usepackage{pgf}
\usepackage{physics}
\usepackage{bbm}
\usepackage{xcolor}
\usepackage{pgfplots}
\pgfplotsset{compat=1.17} 
\usepackage{listings}
\usepackage{qcircuit}
\usepackage{paralist}
\usepackage{hyperref}
\usepackage{float}
\usepackage{afterpage}
\usepackage{microtype}
\usepackage{filecontents}
\usepackage{mathtools}
\usepackage{breakurl}

\usepgfplotslibrary{fillbetween}

\definecolor{codegreen}{rgb}{0,0.6,0}
\definecolor{codegray}{rgb}{0.5,0.5,0.5}
\definecolor{codepurple}{rgb}{0.58,0,0.82}
\definecolor{shadecolor}{named}{lightgray}

\def\colorModel{hsb} 

\newcommand\ColCell[1]{
  \pgfmathparse{#1<0.5?1:0}  
    \ifnum\pgfmathresult=0\relax\color{white}\fi
  \pgfmathsetmacro\compA{0}      
  \pgfmathsetmacro\compB{0} 
  \pgfmathsetmacro\compC{1-#1}      
  \edef\x{\noexpand\centering\noexpand\cellcolor[\colorModel]{\compA,\compB,\compC}}\x #1
  } 
\newcolumntype{E}{>{\collectcell\ColCell}m{0.4cm}<{\endcollectcell}}  

\lstdefinestyle{mystyle}{
    backgroundcolor=\color{lightgray!30!white},   
    commentstyle=\color{codegreen},
    keywordstyle=\color{red!50!orange},
    numberstyle=\tiny\color{codegray},
    stringstyle=\color{codepurple},
    basicstyle=\linespread{0.95}\fontencoding{T1}\footnotesize\fontfamily{lmtt}\fontseries{c}\selectfont,
    breakatwhitespace=false,         
    breaklines=true,                 
    captionpos=b,                    
    keepspaces=true,                 
    numbers=left,                    
    numbersep=5pt,                  
    showspaces=false,                
    showstringspaces=false,
    showtabs=false,                  
    tabsize=2,
    columns=fullflexible,
    language=Python,
}
\usepackage{booktabs,chemformula}
\tikzset{>=latex}
\usepackage{tabularx}
\usepackage[font=small, labelfont=bf, labelsep=period]{caption}

\usetikzlibrary{
    arrows,
    positioning,
    decorations.pathmorphing,
    decorations.markings,
    decorations.pathreplacing,
    shapes,
    fadings,
    calc,
    external
}

\usepackage{algorithm}
\usepackage[noend]{algpseudocode}
\algnewcommand\algorithmicforeach{\textbf{for each}}
\algdef{S}[FOR]{ForEach}[1]{\algorithmicforeach\ #1\ \algorithmicdo}
\algrenewcommand\algorithmicindent{0.55em}%

\def\BibTeX{{\rm B\kern-.05em{\sc i\kern-.025em b}\kern-.08em
    T\kern-.1667em\lower.7ex\hbox{E}\kern-.125emX}}
\begin{document}

\title{Practical Quantum K-Means Clustering: Performance Analysis and Applications in Energy Grid Classification}
\author{
    Stephen DiAdamo$^{1,2}$, Corey O'Meara$^{1}$, Giorgio Cortiana$^{1}$, and Juan Bernab\'e-Moreno$^{1}$\\
    \textit{${}^1$E.ON Digital Technology GmbH}\\
    \textit{${}^2$Technische Universit\"at M\"unchen }\\
    \thanks{
        © 2022 IEEE.  Personal use of this material is permitted.  Permission from IEEE must be obtained for all other uses, in any current or future media, including reprinting/republishing this material for advertising or promotional purposes, creating new collective works, for resale or redistribution to servers or lists, or reuse of any copyrighted component of this work in other works.\newline\newline
        Contact: stephen.diadamo@gmail.com, corey.o'meara@eon.com
    }
}
\maketitle

\thispagestyle{plain}
\pagestyle{plain}

\begin{abstract}
    
In this work, we aim to solve a practical use-case of unsupervised clustering which has applications in predictive maintenance in the energy operations sector using quantum computers. Using only cloud access to quantum computers, we complete a thorough performance analysis of what some current quantum computing systems are capable of for practical applications involving non-trivial mid-to-high dimensional datasets. We first benchmark how well distance estimation can be performed using two different metrics based on the swap-test, using angle and amplitude data embedding. Next, for the clustering performance analysis, we generate sets of synthetic data with varying cluster variance and compare simulation to physical hardware results using the two metrics. From the results of this performance analysis, we propose a general, competitive, and parallelized version of quantum $k$-means clustering to avoid some pitfalls discovered due to noisy hardware and apply the approach to a real energy grid clustering scenario. Using real-world German electricity grid data, we show that the new approach improves the balanced accuracy of the standard quantum $k$-means clustering by $67.8\%$ with respect to the labeling of the classical algorithm.

\end{abstract}


\section{Introduction}

\label{sec:introduction}
Given the challenging engineering requirements for building and maintaining quantum computers, it is likely that quantum computers will only be accessible through cloud services for the majority of users. Quantum computers, depending on the qubit technology, can require a complex construction and maintenance schedule that make it impractical for the average user to own \cite{almudever2017engineering}. Rather than building quantum computers as a hardware product to sell to consumers, companies like Amazon, Microsoft, and IBM are rather developing cloud-based platforms for online access to their quantum devices. Although these quantum cloud services are currently accessible, the question of how useable they are for practical, industrial use-cases arises. For our interest, we ask: Can these quantum computers produce accurate enough distance estimates for clustering? Can we achieve any speedup with them currently? The focus of this work is therefore to benchmark this simple, common use-case using the IBM Quantum cloud services \cite{ibm-services}. 

Clustering algorithms can be used on unlabeled data to find relationships between the data's various features. To perform clustering, an algorithm introduced by Lloyd in 1982 called $k$-means clustering~\cite{lloyd1982least} can be used. The $k$-means algorithm takes as input a collection of unlabeled data points, or feature vectors, and outputs a list of labels, one for each data point. The data points are labeled based on the minimal distance to a particular centroid. During execution, the algorithm improves the centroid locations by running iteratively, updating the location to be the mean of the data points that are determined nearest to them based on a distance metric. 

Classically, the usual method for measuring the distance between centroids and data points is to simply compute the Euclidean distance. For feature vectors of $N$ features, computing the Euclidean distance requires $O(N)$ computational steps. With a quantum approach, using quantum amplitude embedding (see 
\cite{schuld2018supervised} for details), one can encode length-$N$ vectors into $O(\log_2 N)$ qubits, an exponential decrease in resources for embedding, assuming one can load quantum states into a quantum random access memory~\cite{lloyd2013quantum}. With this embedding, one can perform what is known as a swap-test using a quantum computer, as described in~\cite{buhrman2001quantum, lloyd2013quantum}, to compute an estimate for the Euclidean distance between two vectors. Because the swap-test requires a number of operations proportional to the number of qubits used for embedding---needed for swapping two multi-qubit states---in theory this would result in an exponential speedup in runtime complexity. Moreover, the minimum distance to a centroid for each point can be found using a Grover's search \cite{grover1996fast} for an additional quadratic speedup. Using a simpler data embedding approach like angle embedding, the theoretical advantage provided by an amplitude embedding is lost, as angle embedding requires a number of qubits directly proportional to the data dimension. However, the benefit of using this alternative embedding is that the depth of the state preparation circuit is constant, whereas with amplitude embedding, the state preparation circuit uses exponentially more non-local gates as the data dimension grows \cite{plesch2011quantum}, leading to vastly deeper circuits using current approaches. 

This result makes quantum clustering and nearest-neighbor classification appear as very attractive use cases for quantum computing, since they both use distance estimation and therefore can benefit by a theoretical speedup. When put into practice though, there are various challenges to overcome before one can effectively perform distance estimation on a quantum computer. Moreover, a strong assumption of efficient state preparation needs to be made in order to have an exponential speedup using the quantum approach over the classical approach \cite{tang2021quantum}. Nonetheless, with this article, we aim to expand the results related to clustering on real, gate based, quantum hardware, and in particular, to explore how this type of algorithm can be executed on IBM's quantum cloud computing service. In this work, we use only the software libraries and services available to us with no direct hardware access, aiming to demonstrate how well one can expect quantum clustering to perform using generally available resources. 

We begin in Section \ref{sec:setup-config} by reviewing how we perform data encoding and how we calculate the distance estimation in our quantum algorithm using the cloud service. In Section \ref{sec:benchmark}, we benchmark how well distance estimation can be performed, testing for various Euclidean distances and dimensions in simulation and on hardware. Further, we analyze various clustering experiments using synthetic data with two and four dimensions with several datasets. Given the results of the performance analysis in Section \ref{sec:benchmark}, we then apply the findings to a non-trivial clustering problem which is relevant in the energy sector in Section \ref{sec:application}. In particular, we show that decomposing high-dimensional vectors into 2-dimensional subspace projections, we are able to compute the overall distance in a parallel fashion which significantly reduces the error induced by existing quantum hardware, while simultaneously reducing the total number of circuits.

\subsection{Related Work}\label{sec:related-work}

Performing clustering and nearest-neighbor type algorithms using quantum computers has been studied in various contexts. Improving the encoding strategy to work better with IBM's quantum computers was studied by Khan et al. in \cite{khan2019kmeans}. In the article, they describe an encoding mechanism for feature vectors and benchmark the approach on IBM's quantum computer. Feature vectors, after PCA is performed, of dimension two are considered and benchmarked with the MNIST dataset using quantum hardware. Using the IonQ quantum hardware, Johri et al. perform data classification using clustering on their trapped ion quantum computer \cite{johri2021nearest}. In their work, they define an optimized method for encoding 8-dimensional classical data into the quantum computer and use PCA to benchmark against MNIST data for ten different labels and perform their quantum algorithm for nearest-neighbor classification. In these works, an explicit benchmarking of distance estimation accuracy is not demonstrated. Moreover, the only experiments tested on the IBM quantum system were of two dimensions.  In a quantum annealing setting, clustering has also been considered. The authors of \cite{kumar2018quantum} and \cite{arthur2021balanced} map a clustering problem to a quadratic unconstrained binary optimization (QUBO) problem for an adiabatic quantum computer and use hardware to test their approach. Quantum annealing uses a different approach to quantum computing versus the gate based model, indeed no swap-test is involved, and the results from annealing experiments do not paint a clear picture for performance using a universal quantum computing approach.

In \cite{benlamine2019distance}, Benlamine, et al. review three methods for distance estimation using a quantum approach and benchmark the approaches using nearest-neighbor classification in simulation only, not testing their approaches on real hardware. Further modified quantum clustering approaches were proposed in \cite{benlamine2020quantum, qmeans2019kerenidis}. These works did not perform tests on real physical hardware, and therefore did not benchmark their performance, as we do in this work. In \cite{nguyen2021experimental}, Nguyen et al. run experiments to test the accuracy of the swap-test using their trapped ion system in a continuous variable setting, but do not perform any experiments of clustering or distance estimation for classical input vectors. Overall, our work is the first to benchmark these popular swap-test based distance estimation techniques in a general way and for the purpose of classification and clustering.

\subsection{Summary of Contributions}
In this work, we analyze how well current quantum computers can perform unsupervised clustering. To do this, we use two different distance metrics that are based on the swap-test and thoroughly benchmark their accuracy. The swap-test is a general method for computing distance between vectors on a quantum computer and is therefore important to thoroughly understand. The two metrics we use are based on popular, well-known, methods of data embedding, and therefore of a general interest to explore, something this work does for the first time on real quantum hardware. To compare the accuracy of the real quantum device to the true result, we use noiseless simulation as the ground truth. In order to perform many distance measurements at once for $k$-means clustering, we define two approaches to parallelize the calculations. In the first case, we offer a parallelization approach when the quantum computing platform does not allow for sending circuits to execute in bulk. We tested these approaches to perform benchmarking, but because of the large performance improvement when using bulk circuit execution of the second approach, we use only the second approach for our analysis in this work. We benchmark distance estimation via the swap-test, varying the scale of the distance, the number of shots used, and the dimension of the data. To the best of the authors knowledge, no such benchmarking has been previously demonstrated.

Next, we analyze unsupervised clustering on real quantum devices. Using the results of the distance metric analysis, we benchmark the best cases in terms of dimension---2D and 4D---using synthetic data. We found that the accuracy of the real devices using standard approaches and synthetic data proved relatively low. We next tested the ability to perform a true unsupervised clustering problem using real data from the German energy grid. We found that using a standard approach for quantum $k$-means clustering, the accuracy result output from real quantum hardware is low. We then tested our novel approach to performing the distance metrics, which is to decompose the distance calculation into smaller-dimensional projections, and then use quantum circuits that compute multiple swap-tests at once, thereby accommodating for the additional swap-tests required after projecting to smaller dimensions. With this approach, we were able to achieve a high accuracy with respect to the classical results, thus demonstrating a scalable approach to distance estimation with high accuracy on real quantum devices.

\section{Program Setup and Configuration}\label{sec:setup-config} 

In this section, we review the setup and configuration used to perform clustering algorithms. In the first subsection, we review how the synthetic data is generated. Next, we review how the quantum algorithm works as well as the two methods used for classical data embedding in quantum states. Finally, we review the software approach used to execute the algorithms on the quantum cloud hardware.

\subsection{Generating Synthetic Data and Quantum Data Loading}

\begin{figure}[ht]
    \centering
    \includegraphics{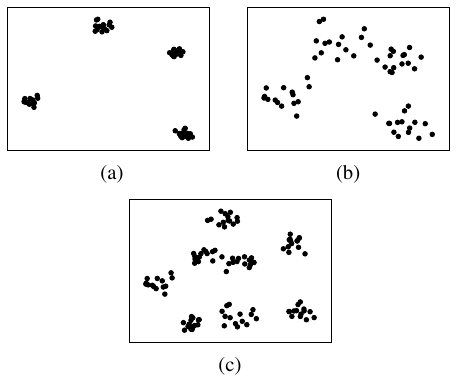}
    \caption{Synthetic 2-dimensional data used for clustering. In (a), the data is tightly clustered with four clusters with 60 points in total which we consider an \textit{easy} clustering. In (b) the data is more scattered with four clusters and 60 points total which we consider a \textit{hard} clustering. In (c) are eight clusters with 14 points per cluster for a total of 112 points.}
    \label{fig:data-samples}
\end{figure}

In the experiments conducted, we used synthetic data generated with varying dimension, number of clusters, cluster variance, and minimal distance between the centers. At a high level, the cluster generation algorithm used works by firstly selecting $k$ center points to then generate cluster data around them. Using the center points as the multi-dimensional mean of a multi-variate normal distribution, a set number of cluster points are generated surrounding the center. Input to this algorithm is a cluster-variance parameter which we use to set the variance level of the respective dimension to control the ``tightness"---how we measure difficultly of clustering---of the cluster. To avoid the randomly initialized center points being too close to each other, an additional step that resets a center point if it is within some $\epsilon$ distance from the already initialized center points is added. The synthetic data generated can be seen in Figure \ref{fig:data-samples}.

\begin{figure}
    \centering
    \includegraphics{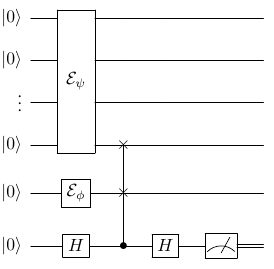}
    \caption{Circuit for embedding classical data using amplitude embedding. We make use of the built in features of \texttt{Qiskit} to initialize the amplitude encoded data.}
    \label{fig:amp-embedding-circuit}
\end{figure}

\begin{figure}
    \centering
    \includegraphics{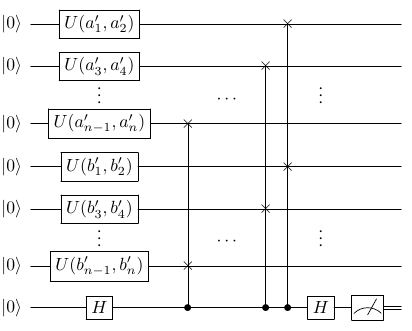}
    \caption{Circuit for embedding classical data using angle embedding with swap test.}
    \label{fig:angle-embedding-circuit}
\end{figure}

\begin{figure}
    \centering
    \includegraphics{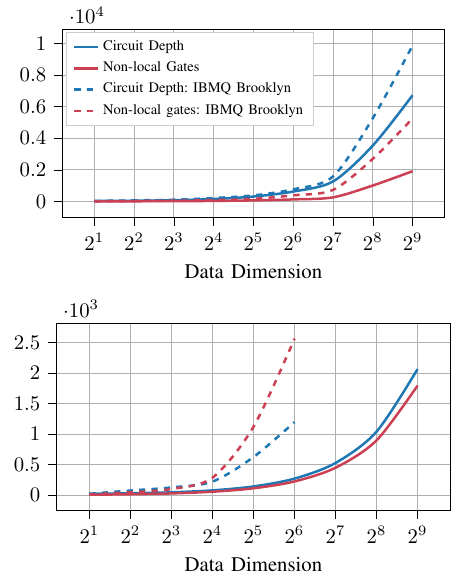}
    \caption{The circuit-depth and number of non-local gates in a circuit that embeds classical data using amplitude embedding and performs the swap test of varying dimension transpiled with the basis gates under full connectivity and for the IBMQ Brooklyn device. The upper plot demonstrates results using amplitude embedding, and the bottom plot from angle embedding.}
    \label{fig:circuit-depth}
\end{figure}

\begin{figure}
    \centering
    \includegraphics{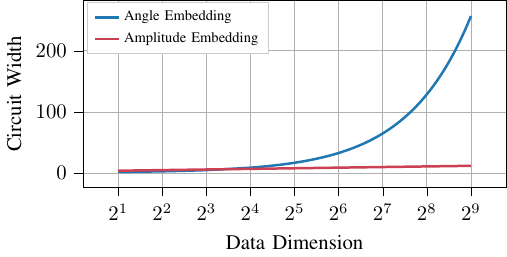}
    \caption{The number of qubits required to embed classical data of varying dimension.}
    \label{fig:qubit-scaling-with-dim}
\end{figure}

To perform the quantum distance estimation algorithm, the generated data points firstly need to be embedded into a multi-qubit quantum state. For this, we use two common types of embeddings, namely amplitude embedding and angle embedding \cite{schuld2018supervised}, and test them independently. To perform the embeddings, we use the circuit structure shown in Fig.~\ref{fig:amp-embedding-circuit} and Fig.~\ref{fig:angle-embedding-circuit} respectively.  To implement amplitude embedding in code, we use the built-in \verb|initialize| function offered by \texttt{Qiskit}, which is a function that takes as input a real vector and returns the necessary gate set for complex amplitude embedding \cite{qiskit-init}. For angle embedding, we use two-dimensional rotations to embed two dimensions of a data vector per one qubit. This type of embedding is also referred to as ``dense angle embedding'' \cite{larose2020robust}, but in this work we will only use the name ``angle embedding'' to refer to it. The scaling in terms of circuit depth and number of two-qubit gates varies significantly between the two embeddings. When embedding data using angle embedding, gate depth does not increase with the data dimension, but circuit width grows linearly. On the other hand, embedding classical data using amplitude embedding can scale quite poorly in depth with respect to the data dimension~\cite{plesch2011quantum}. When using NISQ-era quantum devices, width and depth are some of the properties that should be reduced as far as possible to reduce the effects of noise in any algorithm. We plot the experimental scaling with respect to the data dimension in Fig.~\ref{fig:circuit-depth}. The figure shows both the circuit depth and number of non-local gates required for initializing data for a connected quantum computer (i.e. one in which any two qubits can interact without any state-swapping) (solid line), and the 65-qubit IBM Brooklyn topology (dashed line), followed by a swap-test. For the embedding methods, Fig.~\ref{fig:qubit-scaling-with-dim} shows the qubit resource requirement trends against feature vector dimension. The connectivity of the quantum computer topology determines the number of local swaps needed to perform two qubit gates, which results in deeper circuits with more non-local gates.

\subsection{Quantum Circuits for Distance Estimation}

To perform clustering, we replace the distance calculation from the classical algorithm with a quantum algorithm for distance estimation. Using amplitude embedding, we use a Euclidean distance approximation based on the one developed in \cite{lloyd2013quantum} and reiterated in \cite{kopczyk2018quantum}. For angle embedding, we define a simple encoding that scales the data for embedding. The algorithm used for distance estimation firstly embeds the data for two vectors and then performs a swap-test. To approximate the distance, a number of repetitions, or shots, of the circuit are used to aggregate measurement statistics for a single ancilla qubit. The number of repetitions to use will vary problem to problem and hardware to hardware. Theoretically, using a higher number of shots will produce a more precise distance estimates, but each hardware will have a finite precision, and so more shots does not always lead to more accuracy. Moreover, some datasets do not require high-precision distance estimation to cluster accurately if they, for example, have well separated clusters. We explore this further. Once the repetitions are complete, the distance estimation can be calculated via the probabilities of a 0 or 1 measurement outcome of the ancilla qubit
\begin{align}
    \Pr(0) &= \frac{1}{2} + \frac{1}{2} \left| \braket{\psi}{\phi} \right|^2\;, \\
    \Pr(1) &= 1 - \Pr(0) = \frac{1}{2} - \frac{1}{2} \left| \braket{\psi}{\phi} \right|^2\;, \label{eq:p1}
\end{align}
Thus, the estimate $\left| \braket{\psi}{\phi} \right|^2$ provides an approximation to the inner product for our choice of embedding, or at least a measure that scales with the inner-product in the case of angle embedding.

For each of the two embeddings used, the swap-test procedure differs in terms of number of controlled swaps---also known as Fredkin gates\cite{fredkin1982conservative}---used due to the difference in qubit resources required to perform the data embedding as well as the data representation strategy. For two (not necessarily normalized) data vectors $a:=(a_1, a_2, ..., a_n)$ and $b:=(b_1, b_2, ..., b_n)$---which in the case for clustering or classification would be one centroid and one data point---the two quantum states that are compared when performing the swap-test when using amplitude embedding are given by
\begin{align}
    \ket{\psi} &\coloneqq \frac{1}{\sqrt{2}}\left( \ket{0}\ket{a} + \ket{1}\ket{b} \right)\;,\\
    \ket{\phi} &\coloneqq \frac{1}{\sqrt{Z}}\left( |a|\ket{0} - |b|\ket{1} \right)\;,
\end{align}
where $Z\coloneqq |a|^2 + |b|^2$ \cite{lloyd2013quantum}. To recover the distance estimation in this case, an ancilla qubit initialized in the $\ket{0}$ state is added to the system, and, after applying a Hadamard gate to it, a Fredkin (or controlled-swap) gate is applied with the ancilla as the control with $\ket{\psi}$ and $\ket{\phi}$ as the target systems. In this case, since $\ket\phi$ is just one qubit, one  Fredkin gate is needed, thereby greatly reducing the number of controlled swaps needed. Another Hadamard gate is applied to the ancilla qubit and then it is measured. By repeating the process a number of times, $\Pr(0)$ can be estimated to finally recover the distance: 
\begin{align}
    \left|a - b\right|^2=4Z(\Pr(0) - 0.5)\;.
\end{align}
The circuit for this process is depicted in Fig.~\ref{fig:amp-embedding-circuit}.

For the angle embedding version of distance estimation, we prepare the two angle-encoded data vectors in the following way. Given the classical input vectors $a$ and $b$ defined above, let
\begin{align}
    a'_i &\coloneqq \tfrac{\pi}{2}(a_i + 1) \quad \text{and} \quad \label{eq:a_angle_enc}\\
    b'_i &\coloneqq \tfrac{\pi}{2}(b_i + 1)\;,\label{eq:b_angle_enc}
\end{align}
such that the angle embedded representations are given by $a':=(a'_1, a'_2, ..., a'_n)$ and $b':=(b'_1, b'_2, ..., b'_n)$. The mappings given by Eqns. \eqref{eq:a_angle_enc} and \eqref{eq:b_angle_enc} used to ensure the values are between 0 and $\pi$. With this embedding, we can encode data using a two-dimensional rotation operation defined by
\begin{align}
    U(\theta, \gamma) \coloneqq 
        \left(\begin{tabular}{cc}
            $\cos\tfrac{\theta}{2}$ & $-\sin\tfrac{\theta}{2}$ \\
            $e^{i\gamma} \sin\tfrac{\theta}{2}$ & $e^{i\gamma}\cos\tfrac{\theta}{2}$
        \end{tabular}\right)\;.
\end{align}
We encode the two vectors $a'$ and $b'$ in $\lceil n /2 \rceil$ qubits, using 1 qubit per two dimensions of each vector resulting in the final states
\begin{align}
    \ket\psi &\coloneqq \bigotimes_{i \in odd(n)} U(a_i', a_{i+1}')\ket{0} \quad \text{and} \quad \\
    \ket\phi &\coloneqq \bigotimes_{i \in odd(n)} U(b_i', b_{i+1}')\ket{0}\;,
\end{align}
where $odd(n)$ is the set of odd numbers from $1$ to $n$. To recover the distance estimation, again an ancilla qubit is introduced to the system. A Hadamard gate is applied to the ancilla followed by a series of $n/2$ controlled-swap gates using the ancilla as the control and one qubit from $\ket\psi$ and one from $\ket\phi$. A final Hadamard gate is applied to the ancilla qubit and then is measured. The overall circuit is depicted in Fig.~\ref{fig:angle-embedding-circuit}. The goal in this case is to produce an estimate for $\Pr(1)$, which a valid distance metric. Because we are using arbitrary data which is not normalized in advance, we accommodate for this in the distance metric. For this, we set $Z\coloneqq |a|^2+|b|^2$ such that $a$ and $b$ are normalized as $a=a/\sqrt{Z}$ and $b=b/\sqrt{Z}$. The final distance metric is therefore given by
\begin{align}
    d(a,b) &= \sqrt{Z \cdot P(1)}\;,
\end{align}
where $P(1)$ is given in Eqn. \eqref{eq:p1}. We make the choice not to consider pre-normalized data in this case because when using amplitude embedding with the swap-test, by design there is no need to pre-normalize data. For (dense) angle embedding, a good practice is to consider, based on the dataset, how to divide the interval $[0, \pi)$ into sub-intervals such that each sub-interval has leeway to accommodate over-rotation due to imperfect quantum gates when embedding the data. This would affect the definitions for \eqref{eq:a_angle_enc} and \eqref{eq:b_angle_enc} for each unique dataset we use, as we would need to change the coefficient and the linear shift to accommodate the largest distance between data points in the data set. Therefore, for this work, we do not normalize the data ahead of time to better align with the amplitude embedding approach, and to keep the embedding process general for varying data sets.

\subsection{Running Clustering on the Quantum Cloud}

To implement the circuit preparation and execution of the quantum circuits in software, we use IBM's \texttt{Qiskit} Python Software Development Kit (SDK) \cite{cross2018ibm}. 

A property of the $k$-means clustering algorithm is that it is highly parallelizable, and we use this property to execute the clustering algorithm more time-efficiently. In each iteration of $k$-means clustering, a distance is calculated between the current set of centroids and each point in the dataset. These distance calculations are independent of each other and can be computed in parallel using multi-processing or in a batch job. In the quantum case, to parallelize, we take two approaches. Firstly, we prepare one circuit per distance estimation and then send the collection of circuits as a batch job to the cloud service. The response is the measurement results of the circuits which can then be post-processed according the embedding. The second way we parallelize the algorithm is to embed multiple swap-tests into one circuit and execute them on the same quantum computer at once. This allows for multiple distance estimations to be done on a single quantum computer. 

We developed the first parallel approach in two ways, one using our own approach and later, when it became available, using a \texttt{Qiskit} native approach. For the first approach, we use many computational thread processes (locally) to send requests simultaneously to the server to reduce computation time. In simulation, the approach improved performance, but when there is only one quantum computer, this makes no difference. The advantage from parallel execution is achieved with multiple quantum computers working together to compute distance estimations as described in \cite{parekh2021quantum}. 

The second approach was sending batches of circuit jobs via a recently added feature called \texttt{Qiskit Runtime}. This feature allows the user to submit many circuit jobs at once such that there is only one job to queue. The performance improvement between parallel execution via multi-processing and batch-circuit  execution that we observed was significant. The complete algorithms for the two applied $k$-means clustering approaches are described by Algorithms \ref{algo:process-experiement-multi} and \ref{algo:process-experiement-batch} respectively. 

\textcolor{black}{\begin{algorithm}[ht]
\caption{Clustering on the Quantum Cloud with Multi-processing}
\small{
\textbf{Input:} 
\begin{compactitem}
\item $k$: The number of clusters
\item data: The data to cluster
\item embedding: The choice of data embedding
\item $\epsilon$: The minimum distance between two cluster centers
\item $maxIterations$: The maximum number of iterations to make 
\item $processes$: A list of running processes
\end{compactitem}
\textbf{Output:} An ordered list of labels for the data points
\begin{algorithmic}[1]
    \State $centroids\gets$ initialize the centroids using $\epsilon$ min distance
    \State $convergence \gets$ false, $i\gets 0$ 
    \While{not $convergence$}
        \State $circuits \gets $generate all circuits with the data, centroids, 
        \Statex \hspace{1.6cm} and embedding choice
        \State $dists \gets $ initialize empty shared storage
        \While{not all circuits have been processed}
            \If{a process is idle}
                \State $job\gets$ send single circuit to server and await response
                \State $dist \gets $ process $job$ results according to the embedding
                \State add $dist$ with circuit number to shared storage
            \Else 
                \State wait
            \EndIf
        \EndWhile
        \State Sort $dists$ according to circuit number
        \State $dists \gets$ using the returned, ordered measurement results from 
        \Statex \hspace{1.3cm} the server, complete the distance estimation procedure
        \State $labels \gets $ using the distances to the centroids, label the 
        \Statex \hspace{1.4cm} data points
        \State $centroids \gets $ with the updated labels, recompute the centroids 
        \Statex \hspace{1.95cm} as an average position of the labeled data. 
        \Statex \hspace{1.95cm} Delete centroids for empty clusters.
        \State Check for centroid convergence, update $convergence$ 
        \State $i\gets i + 1$
        \State Check if $i = maxIterations$ and break accordingly
    \EndWhile
    \State return $labels$
\end{algorithmic}}\label{algo:process-experiement-multi}
\end{algorithm}}

\begin{algorithm}[ht]
\caption{Clustering on the Quantum Cloud with Batched Circuits}
\small{
\textbf{Input:} 
\begin{compactitem}
\item $k$: The number of clusters
\item data: The data to cluster
\item embedding: The choice of data embedding
\item $\epsilon$: The minimum distance between two cluster centers
\item $maxIterations$: The maximum number of iterations to make 
\end{compactitem}
\textbf{Output:} An ordered list of labels for the data points
\begin{algorithmic}[1]
    \State $centroids\gets$ initialize the centroids using $\epsilon$ min distance
    \State $convergence \gets$ false, $i\gets 0$ 
    \While{not $convergence$}
        \State $circuits \gets $generate all circuits with the data, centroids, 
        \Statex \hspace{1.6cm} and embedding choice
        \State $job \gets $ send the collection of circuits to the cloud server
        \Statex \hspace{1.05cm} for processing and await response
        \State $dists \gets$ using the returned, ordered measurement results from 
        \Statex \hspace{1.3cm} the server, complete the distance estimation procedure
        \State $labels \gets $ using the distances to the centroids, label the 
        \Statex \hspace{1.4cm} data points
        \State $centroids \gets $ with the updated labels, recompute the centroids 
        \Statex \hspace{1.95cm} as an average position of the labeled data. 
        \Statex \hspace{1.95cm} Delete centroids for empty clusters.
        \State Check for centroid convergence, update $convergence$ 
        \State $i\gets i + 1$
        \State Check if $i = maxIterations$ and break accordingly
    \EndWhile
    \State return $labels$
\end{algorithmic}}\label{algo:process-experiement-batch}
\end{algorithm}

\section{Benchmarking on Quantum Hardware}\label{sec:benchmark}

To develop a clear understanding of how well clustering and nearest-neighbor classification can be performed using current quantum hardware, it is helpful to test how accurately distance estimation can be performed on quantum devices. In this section, we rigorously investigate the performance of distance estimation between vectors of different dimension sizes, number of shots used, standard embedding approaches, and vector distances. Building on this distance measure study, we then benchmark quantum $k$-means clustering which uses the previously described distance calculation by means of low to mid-dimensional synthetic data and clusters.

\subsection{Benchmarking Distance Estimation}\label{sec:dist-benchmark} 

In this subsection, we benchmark a variety of cases for performing distance estimation and compare the estimations using simulation and real quantum hardware. Simulation in this case is done using the noiseless quantum computing simulation platform offered via the \texttt{Qiskit} framework using the OpenQASM backend. Software for the simulation is prepared in exactly the same way in which the quantum circuit logic is sent to the quantum cloud services, we can simply switch the target backend from the simulated backend to the hardware. To perform the analysis, we test three difference cases. We firstly test how varying the number of circuit shots affects the estimation in both two and four dimensions. Next, we vary the distance of the data points also in two and four dimensions. Lastly, we test how varying the data dimensions affects the estimation up to 32 dimensions using amplitude embedding and up to 26 dimensions using angle embedding.

In all tests, for each resulting data point, we use 100 repetitions, plotting the average output and the standard deviation. In each case, we use the measurement error mitigation feature when running the experiments on the physical hardware. To execute the 100 instances, we make 100 copies of the circuits and use the circuit-runner service to execute the circuits in a batched job. 

\begin{figure*}
    \centering
    \includegraphics{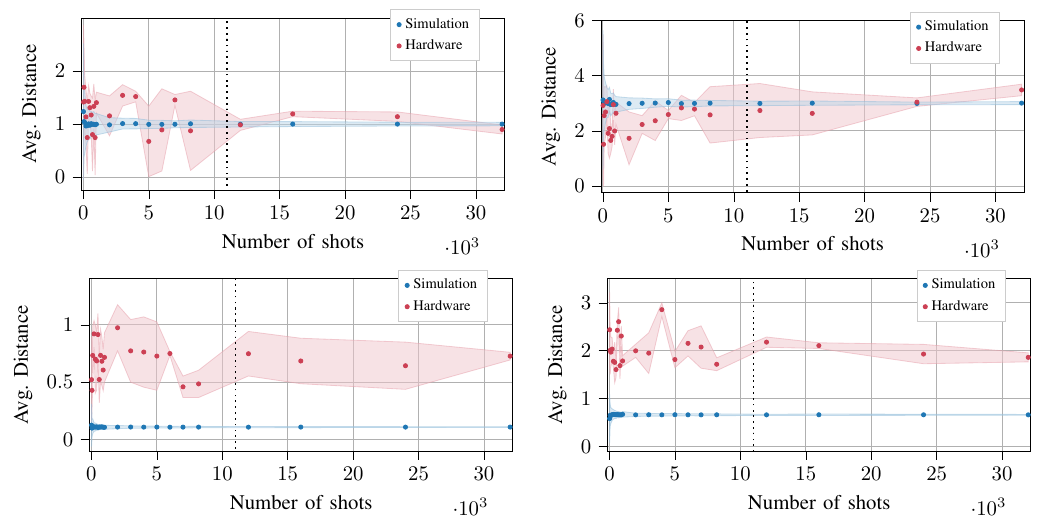}
    \caption{The distance estimation for amplitude embedding (upper) and angle embedding (lower) of $(1, 0)$ and $(1, 1)$ in 2D (left) and $(1,0,0,0)$ and $(1,1,1,1)$ in 4D (right). The circuit for measuring the distance between these the two points is generated and ran a varying number of shots. Displayed is the average output of 100 trials with the standard deviation shaded around the average. The black dotted line indicates where the quantum hardware used switches from IBMQ Bogata on the left of the line to IBMQ Casablanca on the right.}
    \label{fig:est-shots}
    \vspace{5mm}
    \includegraphics{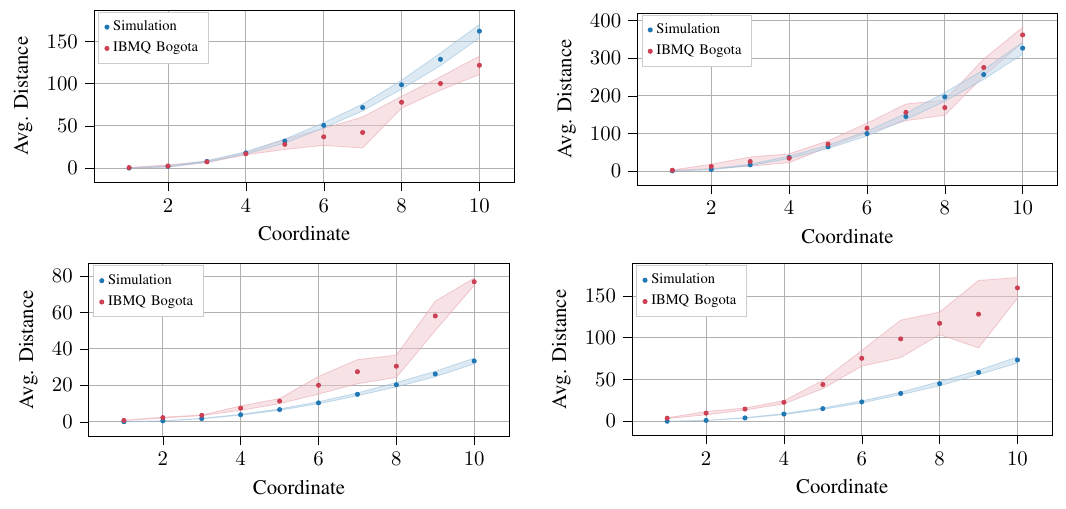}
    \caption{Plots for varying the distance between points. The plots on the left are for amplitude (top) and angle (bottom) embedding in two dimensions. The right plots are for amplitude (top) and angle (bottom) embedding in four dimensions. We run the experiments for a vector of shape $(1, ...,1)$ for the base points and, $(x, ..., x)$ for the varying point. We repeat the experiments 100 times with 2,048 shots plotting the average with the standard deviation.  }
    \label{fig:est-dist}
\end{figure*}

The results of varying the number of circuit shots are plotted in Fig.~\ref{fig:est-shots}. In this experiment, we created a circuit for the vectors $(1,0)$ and $(1,1)$. Comparing simulation to the physical device results, in simulation we see a convergence in the number of shots to the true answer, and moreover the average is close to the true distance as desired. With the real hardware, we see no convergence trends behind the dotted line, and using more shots than around 2,000 does not generally perform better than using the maximum number of shots 8,192. These behaviors indicate that the output probability in the measurements using the real hardware is not consistently Gaussian as one would expect from a noiseless system. Thus convergence between simulation and real hardware to not match in their trends. In some cases, for the same reasons, a lower number of shots performed better, having a lower variance in the standard deviation than with more shots, as seen for example in Fig.~\ref{fig:est-shots}, where the standard deviation is smaller with fewer used shots indicating more consistent outputs overall. Measurement consistency is important for iterative algorithms when many measurements are used, and so a small standard deviation of the estimate is critical. On the other hand, when we switched to a quantum computer that supported a higher number of shots, much better and more consistent results are seen. For 4D, the points $(1,0,0,0)$ and $(1,1,1,1)$ are used, and the same effects are more or less seen. Convergence is not reached using real hardware, but a relatively rapid convergence is seen in simulation, where with a high number of shots, the results improve significantly. We reiterate that in some cases, depending on the dataset, high-precision distance estimation is not necessary for clustering, and therefore this level of accuracy could suffice in some cases.

An important point of note is that for angle embedding, the difference of output between the simulation and the hardware is much starker than with amplitude embedding. The reason behind this is due to the fact that on the IBM quantum devices, the accuracy of qubit rotations is less precise with the available gate basis of $[CX, ID, RZ, SX, X]$ and therefore significant differences between the simulation, where such rotations are highly precise, are observed \cite{finite-precision}. Another noteworthy aspect is that in some instances, the variance in the standard deviation can become very small for the 100 samples, with no recognizable trend. See for example Fig.~\ref{fig:est-shots}, the 7,000 shots point in the upper-left plot and the 6,000 shot point in the lower-left plot. We suspect this may come from a periodic hardware calibration that is performed by IBM that was executed on the quantum devices between experimental runs.

For the next set of experiments, we vary the distance between two points in 2D and 4D (see Fig.~\ref{fig:est-dist}). The vectors we chose have the form $(1,..., 1)$ and $(x, ...,x)$ where we vary $x$ to modify the distance between the points. The results of the 2D experiments show that simulation and the physical device have similar outputs for $x\leq 5$, but for $x>5$, the outputs from simulation and real device start to diverge. In 4D, the effects of embedding show a stark difference between amplitude and angle embeddings. Interestingly, the simulation results for 4D amplitude embedding match very closely to the hardware execution for all tested values of $x$, more so even than in 2D. On the other hand, angle embedding performs far worse in 4D than in 2D, where already for $x = 5$, the difference between simulation and real hardware is significant. This extra noise is again likely due the fact that for more dimensions, the limited precision of the angle embedding is now applied for another two dimensions and that an additional controlled-swap is introduced.

\begin{figure*}
    \centering
    \includegraphics{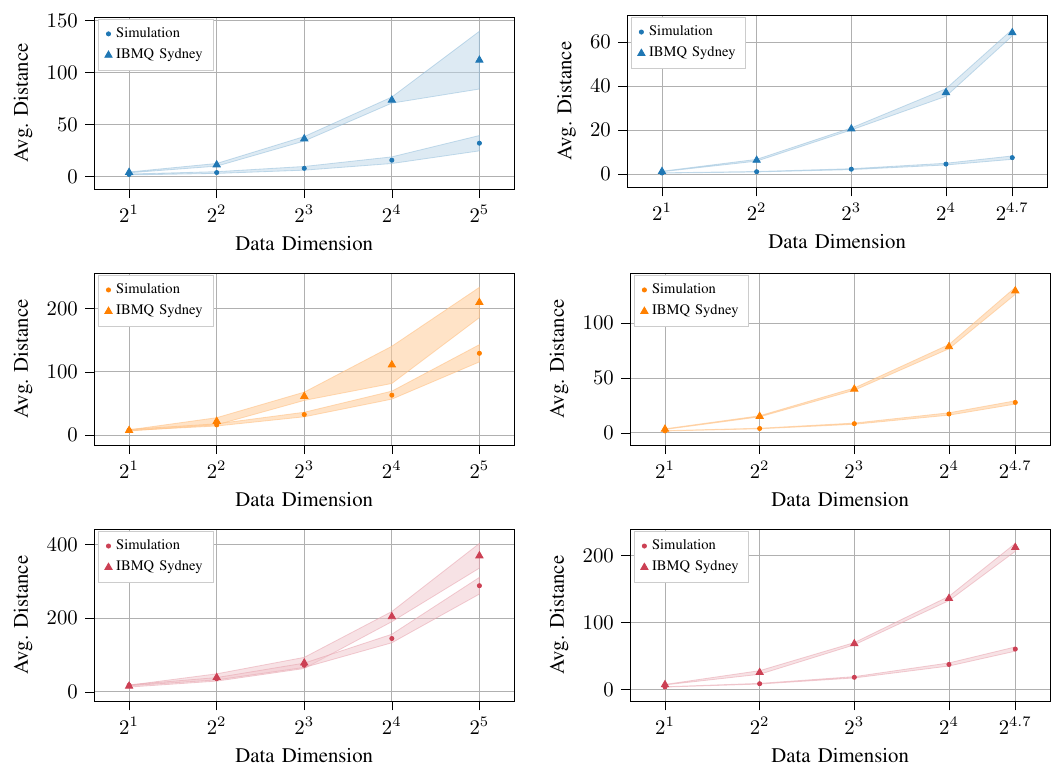}
    \caption{Results for varying data dimension using the vector $(1,...,1)$, estimating distance to vectors $(2, ..., 2), (3, ..., 3), (4, ..., 4)$. The plots are ordered respectively from top to bottom. We run the experiment 100 times with 2,048 circuit shots and plot the average output and standard deviation of simulation after running on the IBMQ Sydney device for up to 32 dimensions with amplitude embedding (left) and 26 dimensions with angle embedding (right).}
    \label{fig:est-dist-dim}
\end{figure*}

In the last set of experiments, we vary the dimension of the data to observe the limits to the number of features we can use and with what level of accuracy. We test up to 32 dimensions for amplitude embedding and up to 26 dimensions for angle embedding, using the 27 qubit device IBMQ Sydney to its capacity in the angle embedding case. We compare vectors with shape $(1,...,1)$, to vectors $(2,...,2)$, $(3,...,3)$, and $(4,...,4)$. In Tables \ref{tab:amp-diffs} and \ref{tab:ang-diffs}, we show the percent-difference between the simulation outputs and the results from the quantum computer for amplitude and angle embedding respectively, where Fig.~\ref{fig:est-dist-dim} displays the corresponding results. Additionally, we show the average gate depth, single-qubit gates, and two qubit gates required to execute the circuit. Circuit transpiling and optimization is a randomized process~\cite{murali2019noise} and so we write the standard deviation for these values after transpiling 100 times. The results demonstrate a clear relationship between total gate depth, 2-qubit gate counts and the data dimensions effect on distance estimation. Although distance estimation using either amplitude or angle embedding approaches each have different gate-depth amounts, we observe empirical evidence of the decoherence effects as a function of problem (data) size. We also observe that at further distances, simulation and hardware results tend to agree more closely in both embedding types. We believe that the accuracy tends to improve as the true distances increase because the normalization factor has a bigger impact on the post-processing, thereby better mitigating inaccurate distance estimates. These results motivate data scaling techniques, where mapping the data into a space where the data is more separated could result in more accuracy, a hypothesis we intend to explore in future work.

\begin{table}[ht]
    \centering
    \scalebox{0.98}{
    \begin{tabular}{c|c|c|c|c|c|c}
        Dim.  & $x=2$ & $x=3$ & $x=4$ & Depth & 1Q & 2Q \\
        \hline\hline
         2 & 66.4\% &  9.6\% & 14.0\% & $25 \pm 1$ &  28 & $12\pm 1$ \\
         \hline
         4 & 97.9\% &  30.3\% & 9.2\% & $37\pm 4$ & 31 & $25 \pm 5$\\
         \hline
         8 & 128.2\% & 61.4\% & 8.4\% & $50 \pm 4$ & 34 & $39\pm5 $\\
         \hline
         16 & 133.2\% & 55.0\% & 34.1\% & $93 \pm 8$ & 37 & $84 \pm 9$ \\
         \hline
         32 & 110.9\% & 47.4\% & 24.7\% & $ 167 \pm 10$ & 40 & $173 \pm 14$ \\
    \end{tabular}}
    \caption{Amplitude Embedding: Percent difference comparison between simulation and real data for the distance between data points of varying dimension and distance on IBMQ Sydney. We write the average circuit depth, number of 1-Qubit gates and number of 2-Qubit gates. Note that, circuit optimization is a randomized process, but with a number of fixed single qubit gates \cite{murali2019noise}, and therefore we show the standard deviation of 100 independent circuit optimizations.}
    \label{tab:amp-diffs}
\end{table}

\begin{table}[ht]
    \centering
    \scalebox{0.96}{
        \begin{tabular}{c|c|c|c|c|c|c}
        Dim.  & $x=2$ & $x=3$ & $x=4$ & Depth & 1Q & 2Q\\
        \hline\hline
         2 & 85.9\% & 61.9\% & 61.3\%  & $25\pm0$ &  27 & $11\pm0$  \\
         \hline
         4 & 144.2\% & 117.9\% & 98.8\% & $53\pm3$ &  47 & $42\pm4$\\
         \hline
         8 & 161.7\% & 131.1\% & 115.9\%& $89\pm7$ &  87 & $99\pm10$ \\
         \hline
         16 & 156.8\% & 128.4\% & 113.8\%& $209\pm17$ & 167 & $364\pm40$ \\
         \hline
         26 & 158.6\% & 129.6\% & 111.4\% & $328\pm22$ &  267 & $611\pm50$\\
    \end{tabular}}
    \caption{Angle Embedding: Percent difference comparison between simulation and real data for the distance between data points of varying dimension and distance on IBMQ Sydney. We write the average circuit depth, number of 1-Qubit gates and number of 2-Qubit gates. Note that, circuit optimization is a randomized process, but with a number of fixed single qubit gates \cite{murali2019noise}, and therefore we show the standard deviation of 100 independent circuit optimizations.}
    \label{tab:ang-diffs}
\end{table}

\subsection{Benchmarking Quantum Clustering}\label{sec:clustering} 

In this section we benchmark the quantum $k$-means clustering algorithm using various dimensions and number of clusters using synthetic data. To determine the accuracy of clustering using the quantum approaches, we generated three types of synthetic data using two and four dimensions. The first two types of data that we generate are an \textit{easy} dataset, a \textit{hard} dataset both with four clusters and 15 data points in each cluster, as seen in Fig.~\ref{fig:data-samples} (a) and (b). This totals 60 data points. The third dataset we use has a variance between the \textit{easy} and \textit{hard} sets, but with eight clusters and 14 points per cluster, seen in Fig.~\ref{fig:data-samples} (c). The number of data points and clusters was selected to most easily work with the \texttt{Circuit Runner} service, reaching the limits to how many circuits can be sent at once. 

\begin{figure*}
    \centering
    \includegraphics{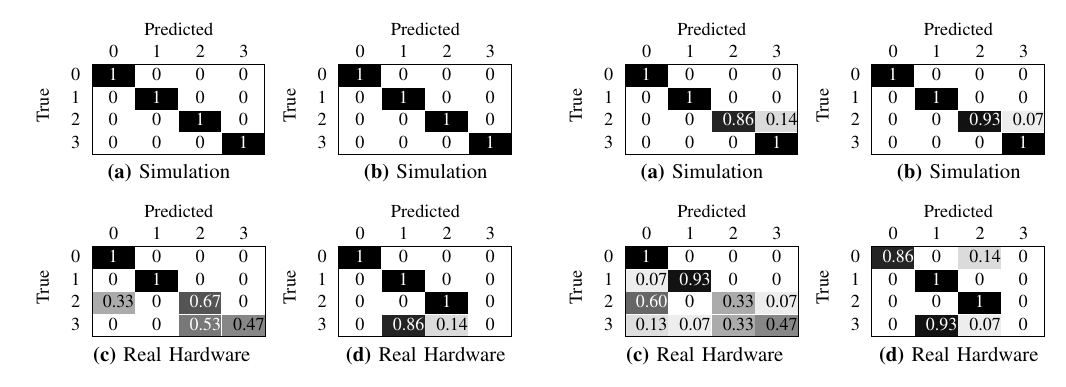}
    \caption{The confusion matrices for the execution for clustering the \textit{easy} (right) and \textit{hard} dataset (left) with two features using amplitude (left column) and angle (right column) embeddings. In (a) and (b) are the clustering outcome of the noiseless quantum simulation.  In (c) and (d) are the results of running on the real IBMQ Sydney device. Each instance ran for a maximum of five iterations using the maximum number of shots 8,192.}
    \label{fig:easy-4}
    \vspace{4mm}
    \includegraphics{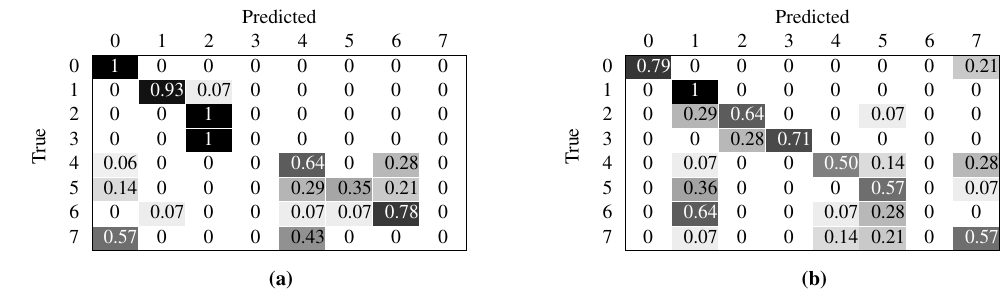}
    \caption{Results from clustering 2D synthetic data as depicted in Fig. \ref{fig:data-samples}(c) using amplitude embedding (a) and angle embedding in (b) on hardware running circuits for 8,192 shots on IBMQ Sydney.}
    \label{fig:hard-8}
\end{figure*}

In the case of the four cluster \textit{easy} and \textit{hard} datasets, we analyze the quantum clustering approach using the same execution parameters for both datasets. We use both amplitude and angle embedding with a maximum of five iterations, or until convergence of the centroid locations is reached. 
We use the outputs of the classical algorithm implementation as the base truth, 8,192 shots for each experiment, and use the option for measurement error mitigation in all cases. For these sets of experiments, we submit jobs to quantum devices which have a quantum volume of 32. The results of the experiments for 2D are seen in Figs.~\ref{fig:easy-4}. The results displayed are, on the left side, left column, the confusion matrices comparing the classical baseline to the quantum outputs using amplitude embedding, and the equivalent for angle embedding in the right column. We observe that in the simulation setting the classical outputs are matched perfectly, however, for real hardware, the results show a relatively low accuracy in the labeling for both embeddings. The results are similar for the \textit{hard} dataset as seen on the right side of Figs.~\ref{fig:easy-4}.

For the 8-cluster data, we perform the same simulation steps, but for a maximum of three iterations with 8,192 shots. In simulation, convergence is reached with two iterations with perfect labeling results. We test both 2D and 4D datasets. The 2D results are shown in  Fig.~\ref{fig:hard-8}. For 4D data the results had very low accuracy, with essentially a random labeling and we neglect showing the results here.

Overall, these experiments motivate that using one instance of executing the distance estimation circuit with low shot counts produce a distance estimation that is not accurate enough to cluster data. To mitigate noise, we predict that we can instead average multiple distance estimations to improve estimation consistency while simultaneously increasing the circuit shots. We test this hypothesis in the next section. 

\subsection{Summary of Benchmarking Results}

In summary, we benchmarked three key points for performing distance estimation and clustering on a quantum computer. Firstly, for distance estimation quality, we benchmarked distance estimation accuracy versus the total number of shots used. We find generally, a high number of shots---larger than 10,000--- should be used to get fairly consistent results. Using values less than 10,000, we find the noise levels are too high to provide high accuracy and consistency. Next, we considered how increasing the Euclidean distance affects accuracy. Here we found that with amplitude embedding, the results trended more similarly to the noiseless, simulated results, whereas the (dense) angle embedding approach tended to diverge from the simulated value. In our last experiment, we investigated how the dimension of the data affects the distance estimation for an increasing dimension size. We found that with amplitude embedding, the simulation results matched more closely to the hardware, but with angle embedding, the values tended to diverge more quickly. We believe this is due to both the additional number of controlled swaps required as well as the finite precision of the physical qubit rotation gates.

Building upon these distance-related metric results, we benchmarked the ability to perform clustering with synthetic data. We tested clustering in terms of varying the dimension of the data, the cluster tightness, and in the number of total clusters. We found that the real hardware performed best in the easiest cases with only 2-dimensional data. By increasing the number of clusters, we expectedly observed a sharp fall in accuracy using our parameter regime.

In the field of quantum computing, being able to produce accurate and consistent results using noisy quantum computers is a major challenge. Here we have conducted an initial study of accuracy, but future work involves finding the best parameter settings such as data dimension and well-clusterable data types that can provide both good performance in terms of runtime and accuracy. In the next section, we propose a scheme that can break down larger-scale problems into smaller problems that are within such a parameter regime.

\section{Applications for Energy Subgrid Clustering}\label{sec:application} 

The motivation for performing a detailed analysis on how well real quantum hardware performs on various distance related metrics for high dimensional data is because, typically, many real-world datasets are non-trivial, high dimensional, and not well clusterable. In classical computing solutions involving unsupervised clustering, the problem of aggregating and grouping sets of objects usually involves some dimension reduction techniques and some type of domain knowledge to tune any machine learning algorithm used. Indeed, a standard $k$-means approach---in the classical sense---when used in practice involves a fine balance of mitigation and optimization techniques depending on the data. Selecting the optimal clusters, choosing the initial centroid locations, choosing an optimal subset size of the dataset for large datasets, dealing with outliers, and dimension reduction are all points to consider when applying $k$-means to an unlabelled dataset. Celebi et al. analyze the effects of these points in \cite{celebi2013comparative}. 

Here, we focus more on the quantum aspect of $k$-means clustering. After using standard techniques for pre-processing the data and selecting initial centroids we direct our efforts to improving the distance estimation accuracy. Since quantum machine learning and variations of clustering are touted as being a possible avenue for quantum advantage, we aim to employ the findings in the previous sections to a real use-case which provides business value. In doing so, we propose an alternative implementation of the distance estimation circuit to overcome some of the deficiencies revealed in the previous section, namely, the inability of the vanilla quantum k-means algorithm to be able to handle classical input vector dimensions more than four.

\subsection{German Electricity Grid Data}\label{subsec:electrical-data} 

Predictive maintenance is a major area of applied research in the energy operations sector \cite{fioravanti2020predictive, mahmoud2021current}. The ability to determine areas of the electrical grid which are susceptible to failing in some pre-determined timespan has many obvious benefits for customers downstream from any grid infrastructure which may fail. One possible approach to this problem is using data-driven analysis of different partitions of the full network grid to group and find similar types of subgrid assets. This can be done by taking into account data features such as 1) the amount of renewable and non-renewable electricity flowing through the grid subsection; 2) the number of power lines within a subsection; and 3) descriptive statistics about the ages of the assets contained in the subsection. Given such a collection of asset properties for electrical grid assets, we aim to employ unsupervised $k$-means clustering to classify various subgrids of part of the German Electrical Grid \cite{carvalho2019systematic}.

The dataset consists of 81,350 low voltage power lines from a Distribution System Operator (DSO) grid in Germany. Each power line has seven numerical features as described in Table~\ref{tab:powerline-data-feats}. Low voltage subgrid networks are connected to high voltage entry and exit points in the grid. For a given high voltage transformer in the grid network, we collect the low voltage lines which are part of its respective sub-network and compute numerical features describing the entire subset of low power lines. Specifically, for each subgrid we compute: the number of non-powerline assets, the total number of connected assets, and the minimum, maximum and sum of each of the features listed in Table~\ref{tab:powerline-data-feats}. There are 1,037 subgrids and therefore we have a final dataset of 1,037 feature vectors, each of dimension 26.

\begin{table}[ht]
    \centering
    \scalebox{1.2}{
    \begin{tabular}{c|c}
        Name & Unit\\
        \hline\hline
        Conductor cross-section & $\text{cm}^2$ \\
         \hline
         Operating Voltage & kV \\
         \hline
         Average Renewable Energy In-feed Load & MWh\\
         \hline
         Average Non-Renewable Energy In-feed Load & MWh \\
         \hline
         Number of Exits of Next Major Substation & \# \\
         \hline
         Line Length & m \\
         \hline
         Sum MVA at closest HV exist & MVA
    \end{tabular}}
    \caption{Description of the features for each powerline in the dataset.}
    \label{tab:powerline-data-feats}
\end{table}

\subsection{Results}\label{subsec:app-results} 

To cluster the 26-dimensional data created using the individual power line features (see Table~\ref{tab:powerline-data-feats}) for all power lines in a given high voltage entry point, we firstly perform a preprocessing step to reduce the total feature vector dimension. In order to fit the data onto the quantum computers available for this work, we reduce the dimension to eight using Principal Component Analysis (PCA) which results in a 97.7\% explained variance as well as a dataset with six dimensions using PCA which accounted for 91.4\% of the variance. This second dataset was used for the angle embedding approach to fit in a 7-qubit quantum computer, the quantum computer topology we had most access to in this work.

\begin{figure}
    \centering
    \includegraphics{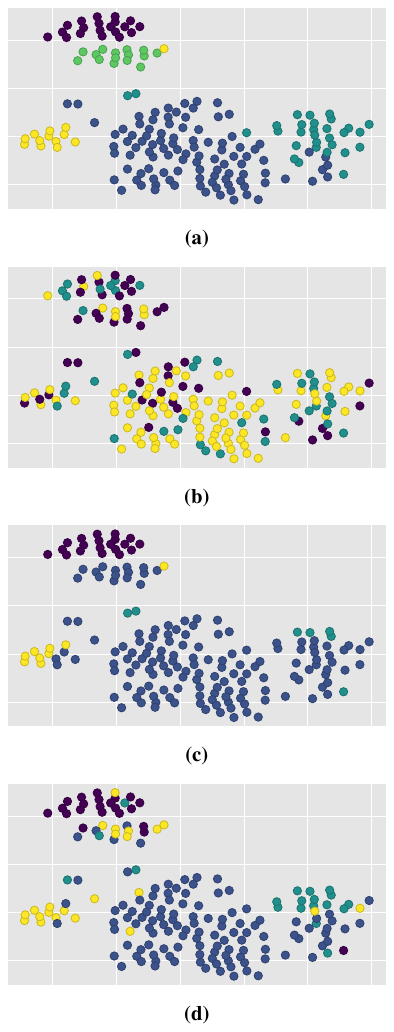}
    \caption{(a) Classical clustering output after t-SNE is performed on the subgrid data. (b) Quantum labeling output using 8D amplitude embedding for five iterations with 12,000 shots per distance estimation on IBM Casablanca. (c) Quantum clustering output using the angle embedding split distance estimation \eqref{eq:metric} of the subgrid data. We decomposed the distance estimation to be one estimate per circuit, for five iterations with 12,000 shots per distance estimation on IBM Perth. (d) Quantum clustering using amplitude embedding split distance estimation \eqref{eq:metric} using a parallel execution process as described (c).}
    \label{fig:tsne}
\end{figure}

With an initial classical analysis using the elbow-method \cite{thorndike1953belongs}, the optimal number of clusters for this dataset was determined to be $k=5$. From this dimension-reduced dataset of 1,087 points, we randomly selected 180 points to cluster, where 180 points allows us to send 900 circuits ($180 \cdot 5 = 900$) to IBM's cloud service in one job (an upper limit for some hardware). Important to any unsupervised clustering algorithm is the choice of initial centroid points. In order to ensure a quick convergence, and to reduce the number of quantum iterations, we ran the classical algorithm with a variety of random seeds such that convergence was reached within three iterations. The classical clustering results are depicted in Fig.~\ref{fig:tsne}(a), using t-Distributed Stochastic Neighbor Embedding (t-SNE) \cite{van2008visualizing} on the high-dimensional data to generate a 2D projection. Using the initial centroids that achieved this, we then ran the quantum clustering experiment. 

To validate the quantum approaches we used, we firstly compare the labeling output from noiseless simulation to the labels output using the classical approach, and then repeat the comparison running on real quantum devices. Because the classical approach converged in three iterations, we allow the quantum versions to run with a maximum of five iterations. In simulation, the balanced accuracy of the experiments were 100\% with amplitude embedding and 97.8\% for angle embedding. Given that simulation produces high-accuracy, we performed a series of tests on the quantum hardware.

\subsubsection{Clustering using amplitude embedding}

The first test we perform is to simply run the same logic as in the simulation. We use 12,000 shots per distance estimate and run the full clustering algorithm for five iterations. The clustering result using amplitude embedding are given in Fig.~\ref{fig:tsne}(b). The grid data is of relative high-dimension and the circuits to prepare the data are roughly 120 gate-depth with approximately 70 non-local gates for amplitude embedding. For angle embedding, the gate depth is expectantly shallower at approximately 86 but with roughly 103 non-local gates, depending on the randomization of circuit transpilation step. With the level of noise occurring, five iterations does not improve the results, and indeed we speculate further iterations would not have led to improved results either. Here, our observation is that the labeling is essentially random due to the noise in the distance estimation circuits, never leading to a converging state.

\subsubsection{Classification using amplitude embedding}

\begin{figure}
    \centering
    \includegraphics{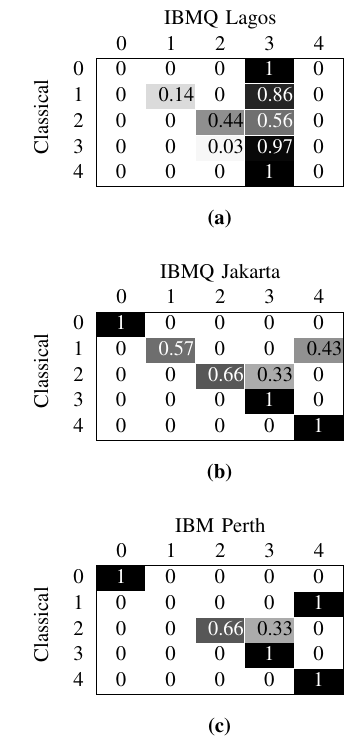}
    \caption{The results of classifying the test set of 60 data points using a nearest-neighbor prediction. In (a), we use amplitude embedding in 8 dimensions and 30,000 shots per distance estimation. The balanced accuracy in this case is 17.0\% and the raw accuracy 46.7\%. Weighted precision cannot be computed since some classes are empty. In (b) we repeat the classification using 15,000 shots using the divided distance estimation in four 2D estimates, using an averaged five estimates per distance. The results have a balanced accuracy of 84.8\%, a raw accuracy of 90.0\% and a weighted precision score of 92.1\%. In (c), the classification is done using the vector subspace parallelization circuit which uses two swap-tests per circuit. This approach reduces the total number of circuits by 50\%. The results improved significantly to have a balanced accuracy of 73.3\% and a raw accuracy of 83.3\%.} 
    \label{fig:grid-classification}
\end{figure}

As a second test, we implemented a pure nearest-neighbor classification application. We begin by training the model offline classically to determine optimal centroid locations, then, at runtime, we compute only the prediction step quantumly to determine which cluster test set data points belong to. Fig.~\ref{fig:grid-classification}(a) shows the accuracy results of the outcome, where we used 30,000 shots to estimate the distances using amplitude embedding in 8D. We see the majority of points were assigned to one class, similarly to how the five iterations of clustering performed.

\subsubsection{Distance estimation with vector subspace parallelization}

So far, our approach to distance estimation has been to use the approach as stated in \cite{lloyd2013quantum}, however, the accuracy in practice using this approach has thus far been relatively low. From the benchmarking section, the highest accuracy was seen in the two-dimensional data experiments. Using this as motivation, we propose a new technique of parallelizing the distance calculation for high dimensional vectors by using distances between two-dimensional subspaces of the full feature vectors. In classical clustering problems, for large dimensional data sets, subspace clustering has been considered to cluster subspaces of the data \cite{parsons2004subspace}. In our case, we can use approaches for large dimensional datasets, but for small datasets, accommodating near-term quantum computers. We consider the entire space as a sum of the subspaces, rather than to find clusters in any particular subspaces, as described in \cite{parsons2004subspace}.

\begin{figure}
    \centering
    \includegraphics{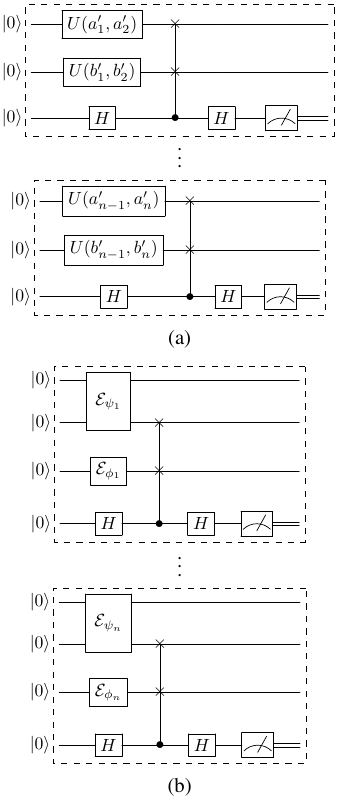}
    \caption{The circuit for processing multiple distances in one circuit using (a) angle embedding and (b) amplitude embedding with swap tests. The dashed boxes separation each independent swap-test.}
    \label{fig:double-angle-embedding-circuit}
\end{figure}

Given input data vectors $a\coloneqq(a_1, a_2, ..., a_n)$ and $b\coloneqq(b_1, b_2, ..., b_n)$, the distance between them can be decomposed as
\begin{align}\label{eq:metric}
    \begin{aligned}
    d(a, b) = \hspace{1mm}&d(a_{1,2}, b_{1,2}) + d(a_{3,4}, b_{3,4})  \\ &+  \dots +         d(a_{n-1,n}, b_{n-1,n}),
    \end{aligned}
\end{align}
where $a_{i,j}=\mathcal{P}_{i,j}(a)$ and $b_{i,j}=\mathcal{P}_{i,j}(b)$ are projections of the respective vectors to the $(i,j)$-th vector subspace. The circuit for this parallel distance estimate using angle embedding is depicted in Fig.~\ref{fig:double-angle-embedding-circuit}(a) and amplitude embedding in Fig.~\ref{fig:double-angle-embedding-circuit}(b). One can extend this approach to any form of 2D data embedding, filling a quantum processor with many simultaneous distance estimates. Moreover, as quantum computers improve, more dimensions can be included in the decomposition, for example, using 4D decompositions instead of 2D. When filling multiple circuits into one quantum processor, we need to be concerned about the cross-talk \cite{ohkura2022simultaneous}, and future work will be to investigate optimal qubit layouts to maximize performance.

This approach has various benefits in terms of mitigating noise. Firstly, it uses only low-dimensional projections. In this case, we use two-dimensional projections, aligned with our benchmarking results, but as hardware improves, we can extend this to larger dimensions to reduce the number of total independent measurements until we can eventually use the entire vector. Next, these low-dimensional circuits will be in general shallower and thinner, which will improve the accuracy and moreover reduce the computation time, allowing for more shots within the same execution timespan. Because in some cases we observed a large standard deviation, with shorter execution time, one can also execute the circuit many times to produce an average distance estimate the same timespan, mitigating Gaussian noise in the system.

Because the distance estimation circuits are thinner, we can load multiple circuits into one quantum processor proportional to the number of qubits. For example, because the swap-test with angle embedding in 2D uses three qubits per swap-test, we can load two distance calculations at a time in a seven qubit quantum computer, reducing the number of total circuits to execute by 50\%. This approach could be generalized to contain as many swap-tests as there are (the floor of) one third the number of qubits, which could in turn result in again using one circuit for distance estimation, simply with a modified pre- and post-processing step. These benefits make this much more NISQ compatible than performing the distance estimation with all dimensions considered at once.  

After verifying this approach produced accurate results in simulation, we tested how well it mitigates the effects of noise in the classification task, we ran the circuit implementing \eqref{eq:metric} using two approaches. For the first approach we used amplitude embedding and executed each distance estimation circuit independently five times, averaging the results and using the average as the distance estimate. We used 15,000 shots per execution and since the circuit uses four qubits, we could fit just one circuit at a time on the 7-qubit device. The confusion matrix of the results is in Fig.~\ref{fig:grid-classification}(b), showing a vast improvement. The balanced accuracy in this case is 84.8\% compared to 17\% using the standard quantum k-means algorithm, an improvement of 67.8\%.

For the second approach, we perform classification again, now using angle embedding, but in this case, since just three qubits per swap-test are required, we loaded two distance estimates in parallel into the 7-qubit device. Using 15,000 shots with, in this case, two repetitions per circuit, we again use the average for the estimate. The results of the classification are shown in Fig.~\ref{fig:grid-classification}(c). Again, we see a strong improvement for the classification problem over the standard method of embedding all the data at once and using a larger number of controlled swaps, increasing the balanced accuracy from 17\% in the standard approach to 73.3\% using this novel approach, a difference of 56.3\%.

\subsubsection{Clustering with vector subspace parallelization}
Given the promising results from the classification task using the vector subspace parallelization, we again perform the full clustering algorithm using the angle embedding approach. We  use the distance estimation \eqref{eq:metric} with 18,000 shots and one measurement repetition on the IBM Perth machine with a total of five iterations for the clustering algorithm. The clustering results are shown in Fig.~\ref{fig:tsne}(c). Although the labels were reduced to four classes, one fewer than in the classical algorithm, we see a much clearer separation of the classes in comparison to using amplitude embedding processing all eight dimensions in the standard approach. The balanced accuracy compared to the classical label results from clustering in this case are 58.5\%. We use the same subspace approach with 8-qubits on the IBM Montreal machine, using two amplitude encoded circuits as in Fig.~\ref{fig:tsne}(d) with 12,000 shots and a balanced accuracy of 63.9\% in comparison to the classical clustering label results. Overall, these results show a vast improvement to the standard approach which would consider all dimensions in one measurement.

\section{Conclusion and Outlook}\label{sec:conclusions} 

In this work, we thoroughly investigated the potential of using quantum $k$-means clustering in a practical manner on current NISQ quantum hardware. In terms of distance estimation comparison between classical and quantum distance calculations, we clearly observed a high level of difference between simulation and running on physical devices---especially comparing the distance estimation results using angle embeddings. The results which used batched job submission via \texttt{Qiskit Runtime} showed to vastly improve performance, allowing for more circuit executions, improving reliability, as well as drastic speed improvements when dealing with large datasets due to the reduction in job-queuing time. 

The best $k$-means clustering results observed were from clustering datasets of 2-dimensional data points. When we increased the number of clusters from four to a more complex scenario of eight clusters, and changed the input vector dimension from two to four, the results worsened. We experimented with an industrial unsupervised learning problem, labeling high-dimensional energy grid data using $k$-means clustering. Using the standard approaches, the clustering and classification results proved inaccurate when executed over real hardware. When we changed the distance metric and used our vector subspace parallelization approach, we saw a significant improvement in both our classification and clustering experiments. For amplitude embeddings, the balanced accuracy of the classification went from 17\% using the standard approach to 84.8\% with this novel distance estimation method. With angle embedding, loading two swap-tests into one circuit to execute in parallel, albeit an overall wider circuit, proved to also have a large performance improvement over the amplitude embedding approach, with a balanced classification accuracy of 73.3\% and raw accuracy of 83.3\%. 

This work provides a first step into quantum clustering for practical, industrial use-cases, but still there are questions to be answered. Future work will be to consider other clustering algorithms such as $k$-medoids which uses alternative distance metrics, considering other quantum approaches for distance estimation as in \cite{benlamine2019distance, kuzmak2021measuring} or those better suited for NISQ hardware as proposed in \cite{fanizza2020beyond}. Indeed, many algorithms require a distance calculation step, and so benchmarking their quantum performance leaves many possibilities for future work.

Although it is well known that quantum computing is in its early stages of development, it is important to investigate what boundaries exist in relation to non-trivial problems that move beyond fundamental algorithm proof-of-concepts. Clustering, and particularly distance estimation, are widely used in various industry applications. With this work, we have tested a large set of experiments that can be performed on the quantum cloud using only the core features of the platform. Even though we do not believe in the current stage, using quantum computers for clustering will add a benefit, quantum technology is continuously and rapidly improving. With this, we expect that as NISQ-era quantum computers mature, these types of analysis and industry-driven use-case studies will continue to provide valuable insight into how they will be used for real-life applications.

\section*{Acknowledgment}

We gratefully acknowledge Arthur Kosmala for his contributions to running portions of the parallelized distance metric experiments used in the Energy Grid use-case. We also acknowledge the use of IBM Quantum services for this work. The views expressed are those of the authors, and do not reflect the official policy or position of IBM or IBM Quantum. S.D. completed this research as part of the Munich Quantum Valley, supported by the Bavarian state government with funds from the Hightech Agenda Bayern Plus and was supported partly via the Emmy-Noether grant no. 1129/2-1 of the German Research Foundation (DFG, Deutsche Forschungsgemeinschaft).

\bibliographystyle{IEEEtran}
\input{refs.bbl}

\end{document}